# Dynamic Control of Third-order Nonlinear Optical Properties of Gold Nanoparticle/Liquid Crystal Composites under External Electric Fields

Shengwei Wang, Mohamed Amine Gharbi,* Chandra S Yelleswarapu*

Department of Physics, University of Massachusetts Boston, 100 Morrissey Blvd, Boston, MA 02125.
*MG (Mohamed.Gharbi@umb.edu; Tel: 617-287-5997);
*CY (Chandra.Yelleswarapu@umb.edu; Tel: 617-287-6063).

This study investigates the dynamic control of third-order nonlinear optical absorption properties of gold nanoparticles (AuNPs) dispersed in nematic liquid crystals (LC). By leveraging the reconfigurable nature of liquid crystals under external electric fields, we demonstrate the ability to manipulate AuNP alignment, dimer formation, and subsequently the plasmon coupling effects. Planar oriented and degenerate LC cells were prepared, and their optical responses under varying electric fields were characterized using polarization microscopy, UV-VIS spectroscopy, and Z-scan techniques. In planar cells, the applied electric field reorients LC molecules and AuNPs, influencing plasmon coupling and the nonlinear absorption. Conversely, degenerate cells exhibit more complex behaviors due to multiple LC alignment directions. These findings illustrate the potential of AuNP/nematic LC systems for creating tunable photonic devices responsive to external stimuli.

## 1 Introduction

The reconfigurable nature of liquid crystals (LCs) has captivated researchers for decades due to their unique properties of fluidity and order, which can be manipulated by external stimuli such as electric fields, magnetic fields, or light.[1–3] This adaptability makes LCs exceptionally versatile materials with a wide array of potential applications.[4,5] In addition, LC orientation can be controlled using patterned substrates, wherein the substrate is etched with microscopic structures like grooves or ridges, inducing a preferred alignment of the LC directors. By combining external stimuli with patterned substrates, the LC system may find applications in optical modulators, switches, and tunable filters. Another important characteristic of LCs is their topological defects, which are regions where the director is ill-defined. These defects are easy to create and manipulate. They have been widely used to organize colloidal dispersions and to template polymerization, leading to a variety of assemblies useful for various applications.[6,7]

On the other hand, the field of nonlinear optics has been instrumental in driving advancements in photonic technologies, facilitating a wide array of applications. The interaction between material and intense light results in third-order nonlinear optical effects, such as nonlinear refraction and absorption. Research over the past several decades has focused on the design and synthesis of a diverse range of materials with the objective of improving the third-order nonlinear optical coefficients. This includes organic molecules and polymers, electro-optic polymers, and metal nanoparticles.[8–18] Among the metal nanoparticles, gold nanoparticles (AuNPs) are particularly noteworthy due to their chemical stability, biocompatibility, ease of synthesis, and localized surface plasmon resonance (LSPR) response.[19–21]

LSPR is a phenomenon that occurs when conduction electrons on the surface of metallic nanoparticles resonate with incident light at specific wavelengths. This resonance leads to a strong enhancement of the local electromagnetic field around the nanoparticles, resulting in significant linear and nonlinear optical effects.[22–29] Various methods have been employed to fabricate AuNPs with specific sizes and shapes, doping concentrations in dielectric medium matrices, and synthesize composites with other materials.[30–33] Cesca et al. demonstrated that by changing the ratio of gold (Au) to silver (Ag) in nanoprisms, the material's nonlinear absorption properties could be modified.[34] Additionally, the same group created a unique Au-Ag nanoplanet configuration by ion implantation and irradiation in silica, resulting in tunable absorption under variable intensity.[35] Hua et al. investigated the nonlinear optical properties of differently shaped AuNPs, showing that nanorods and nanostars exhibit relatively stronger responses.[36]

Dispersing nanoparticles in LCs provides fertile ground for developing novel materials with unique, synergistic properties.[37–42] When metallic nanoparticles are integrated into LCs, the resulting nanocomposite materials exhibit remarkable electro-optical behaviors.[43–50] For instance, we recently demonstrated enhanced nonlinear absorption by aligning gold nanospheres (AuNPs) in nematic LCs.[51,52] When AuNPs are dispersed in 5CB (4-Cyano-4'-pentylbiphenyl), elastic forces cause dispersed AuNPs to align parallel to the 5CB director to minimize elastic distortions brought about by the presence of particles. Thus, AuNPs follows the long axes of the LC molecules, forming complex structures such as nano dimers. Compared to a single nanosphere, these complex structures have shown strong enhancement in the localized electric field, leading to further enhancement of nonlinear optical effects.

Several studies have shown that applying an external electric field can effectively modulate the nonlinear properties of materials.[5,38,53–59] For instance, Mbarak et al. investigated the nonlinear refractive index of AuNPs doped with NLC and found that it could be controlled by varying compositional percentages and applying an electric field.[60] Cantillo et al. characterized nonlinear absorption and refraction of gold nanorods in Cargille oil and observed an increase in the refraction coefficient and a decrease in absorption with increasing electric field strength.[61] Koushik et al. exploited the Kerr effect to control the nonlinear optical properties of aluminum-doped zinc oxide nanoparticles. The nonlinear refractive index switched from a negative to a positive value as the applied voltage was increased from 160 V to 260 V. A corresponding behavior was observed in the nonlinear absorption coefficient.[62] These studies suggest that an electric field can be used to tune the nonlinear optical response of materials, which could have potential applications in nonlinear optics, optoelectronics, and other fields.[63]

Traditional materials, however, often suffer from limited tunability and fixed optical properties, restricting their adaptability in dynamic photonic systems. In this study, we explore the dynamic controllability of AuNPs) dispersed in nematic LCs. By dispersing AuNPs in a 5CB nematic LCs, we created a reconfigurable material whose third order nonlinear optical absorption can be controlled by an external electric field. We demonstrate that applying a low external electric field can orient the formation of AuNP dimers by reorienting the LC molecules. This reorientation affects the plasmon coupling and enhances the third-order nonlinear absorption properties of the material. The dynamic photonic properties of these AuNPs/5CB systems offer potential for creating novel tunable optical devices that can be actively controlled in real-time using external electric fields.

## 2 Experimental Section

Linear and nonlinear optical characterization is performed by applying 1 kHz square wave field with peak-to-peak voltage ranging from 0 V to 10 V is applied across the LC cell.

### 2.1 Sample preparation

The AuNPs solutions with 20 nm in diameter with a plasmon resonance at 520 nm were purchased from Nanopartz. To incorporate them into 5CB nematic LCs, DI water was evaporated from the solution and then were redissolved in ethanol. Then 5CB was introduced to mixture and was heated while stirring until ethanol is completely evaporated. The resulting AuNPs/5CB mixture (the concentrations of both are 0.05 wt%) was injected into prefabricated indium tin oxide (ITO) coated glass cell. As shown in the Figure 1, two ITO coated glass slides (25 x 25 x 1 mm, SIGMA-ALDRICH) were cleaned with sulfuric acid and then cleaned with water. The slides' ITO surfaces were coated with a monolayer of polyvinyl alcohol (PVA) solution (90% water, 10% ethanol, 3% PVA). The slides were placed in a 110 °C oven for 1 hour to achieve planar anchoring by the PVA film. For planar-oriented anchoring of 5CB molecules, the PVA film is mechanically rubbed with a soft cloth in a specific direction to induce microscopic grooves and creates uniform director field, which is essential for achieving anisotropy in the sample. A 20 μm Mylar spacer (Grafix wrap) was placed between the two slides, and the cell was sealed with optical glue. To prepare planar-degenerate samples, a similar procedure was followed except rubbing the surface with cloth. Two copper tapes were attached to the ITO coated surfaces, soldered with wires, and connected to a function generator. The electric field is generated through the AuNPs/5CB was perpendicular to the two slides.

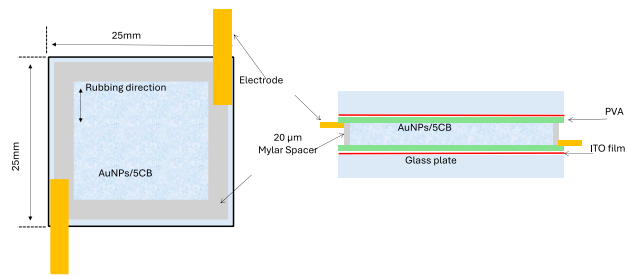

Figure 1 Schematic of fabricated LC cells. The left panel shows the top view, with AuNPs dispersed in 5CB LC. The rubbing direction is indicated by arrows to show the alignment of the LCs. The right panel shows a cross-sectional view of the cell, consisting of two glass plates, ITO and PVA layers, Mylar spacer, AuNPs/5CB mixture in between, and electrodes to applied electric field across the LC cell.

### 2.2 Z-scan technology

Schematic of the open aperture Z-scan system is shown in the Figure 2. The laser system uses a frequency-doubled Nd:YAG laser (Continuum Minilite II) emitting at 532 nm with a pulse width of 3-5 ns and a beam waist of 55 µm. A 20 mm biconvex lens focuses the laser beam, maintaining consistent intensity for all Z-scan measurements. The AuNPs/5CB sample, mounted on a translation stage (Thorlabs NRT 150), connecting with a function generator, moves along the z-axis through the laser beam's focal point. Transmittance data are collected using an optical detector (Newport 818-SL), and the stage translation and data acquisition are controlled through a LabView routine. The output power of the laser for which the all the Z-scan experiments were performed in 110 µJ.

From the recorded data, a normalized transmittance curve was obtained, allowing for the determination of the nonlinear absorption coefficient β, using the following equation:[64]

$$T(Z) = \sum_{m=0}^{m=\infty} \frac{\left(-\beta \frac{I_0}{\left(1+\frac{z^2}{z_0^2}\right)} L_{eff}\right)^m}{(m+1)^{\frac{3}{2}}} \quad L_{eff} = \frac{(1-e^{-\alpha_0 L})}{\alpha_0} \qquad (1)$$

where $I_0$ is on axis light intensity. The Rayleigh range $z_0$ is given by equation $\frac{\pi \omega_0^2}{\lambda}$, where $\omega_0$ and $\lambda$ are beam waist and wavelength of laser source, respectively. The linear absorption coefficient $\alpha_0$ is calculated from absorbance values using Beer-Lambert law. The effective length $L_{eff}$ is related to $\alpha_0$ and L, the sample thickness.

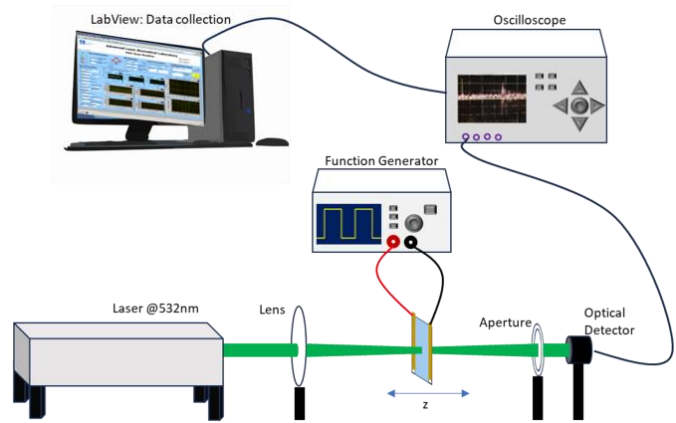

Figure 2 Experimental setup schematic for measuring the nonlinear optical response of the AuNPs/5CB LC cell. A nanosecond Nd:YAG laser pulses are focused by a lens. The LC cell is scanned through the focal region (indicated by the bidirectional arrow) and the transmitted light is then detected by a photodetector and analyzed using an oscilloscope and LabView. The function generator provides 1 kHz square wave signal, which is applied across the LC cell.

## 3 Results and discussion

LCs are sensitive to electric fields. An external electric field can orient the 5CB director axis in the direction of the applied field, as shown in the Figure 3. In the case of planar oriented LC cell, at 0 V or no field is applied, the 5CB molecules are aligned along the induced microscopic groves due to the strong anchoring effect at the interface between the LCs and the rubbed substrate. Additionally, intermolecular forces between the 5CB molecules themselves contribute to this alignment. These forces, primarily van der Waals interactions, cause the molecules to align parallel to each other, promoting uniformity across the entire cell. Together, these effects ensure that the nematic LCs uniformly align along the direction of rubbing, maintaining the desired orientation from one end of the cell to the other, as depicted in Figure 3(a).

When the applied voltage across the LC cell is less than the critical voltage, Vc, (the voltage needed to reorient the molecules significantly), the alignment is only minimally disturbed because the voltage is below the critical level and the anchoring forces at the surface are strong enough to maintain the alignment along the direction of rubbing. However, the applied electric field induces a small torque on the 5CB molecules in the interior layers, causing a slight reorientation in the direction of the applied field, as illustrated in Figure 3(b).

When the applied voltage exceeds the critical voltage, the electric field becomes strong enough to overcome the anchoring forces at the surfaces and the intermolecular forces between the 5CB molecules. As a result, they reorient themselves in the direction of the applied field, as shown in Figure 3(c). This reorientation changes the optical properties of the LC cell, allowing for dynamic control of light passing through the cell.

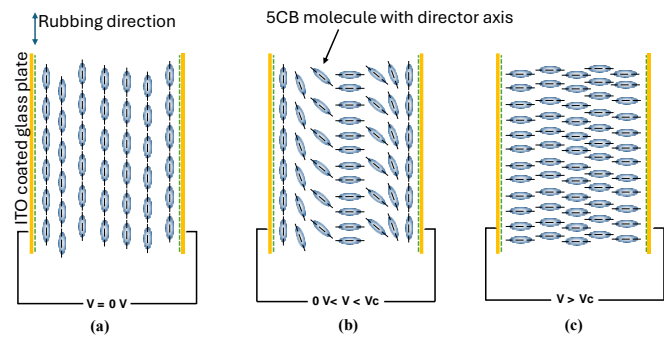

Figure 3 Schematic illustrating the dynamic control of liquid crystal molecules when electric field is applied across planar oriented cell, when applied voltage is a) 0 V, b) below critical voltage, and c) above critical voltage.

In a degenerate LC cell, where the substrates do not impose a single preferred direction, the behavior of the LCs can be more complex. When no voltage is applied, the LC molecules still experience some degree of alignment due to complex anchoring at the surfaces and intermolecular forces, but the molecules orient along any of the equivalent directions, resulting in a more random or multidomain alignment. When the applied voltage is less than the critical voltage, the LCs experience slight reorientation and tend to follow the applied electric field direction to a small extent, but the alignment remains relatively disordered. When the applied voltage exceeds the critical voltage, the electric field's influence becomes dominant. The LC molecules reorient to align uniformly in the direction of the electric field, significantly altering the cell's optical properties. As in planar oriented

LC cell, this reorientation is reversible: once the voltage is removed, the LCs gradually return to their initial, more disordered state, guided by the residual anchoring effects and intermolecular interactions. This behavior allows for the dynamic modulation of optical properties, although with less precision and uniformity.

When a AuNPs/5CB mixture is introduced into oriented and degenerate LC cells, the interaction between the nanoparticles and the LC molecules is significantly influenced by the elastic forces present in the LC. In the oriented LC cell, the elastic forces cause the nanoparticles to align parallel to the 5CB director. This alignment minimizes elastic distortions that would otherwise be introduced by the presence of the nanoparticles, ensuring that the LC molecules remain uniformly oriented, as shown in figure 4. In a degenerate LC cell, where the alignment layers do not impose a single preferred direction but allow multiple equivalent orientations, the nanoparticles tend to align with the local 5CB director, despite the multidirectional potential alignments of the LC molecules. This alignment reduces the elastic distortions and helps maintain a relatively stable configuration within the degenerate cell.

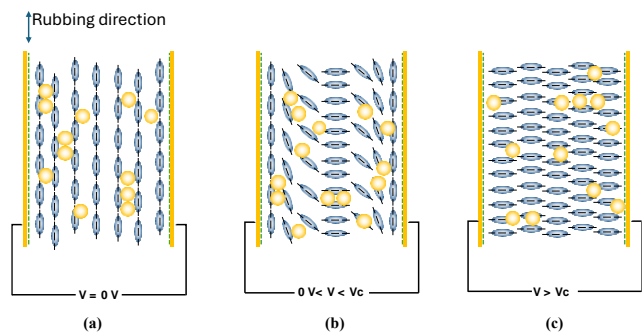

Figure 4 Schematic illustrating the dynamic control of AuNP/5CB mixture under the influence of an electric field applied across a planar-oriented cell, when applied voltage (a) is 0V, (b) below the critical voltage, and (c) above the critical voltage.

### 3.1 Microscopic images

The alignment and orientation of the AuNPs/5CB samples were studied using polarizing microscopy (AmScope, 50x objective). In the cross-polarized configuration, the transmission axes of the polarizer (P) and analyser (A) are crossed (90° to each other). For the study of the oriented LC cell, the rubbing direction is parallel to P. When no voltage is applied across the LC cell (V = 0 V), no light passes through the cell because the LC directors align with the rubbing direction and does not affect the polarization of the light; the light is effectively blocked by the crossed polarizers. However, when the crossed polarizers are rotated 45° relative to the easy axis of the nematic cell, some components of the transmitted light are allowed to pass through, resulting in observable brightness in the samples (Figure 5-a). These observations demonstrate the effective control of anchoring properties in our samples.

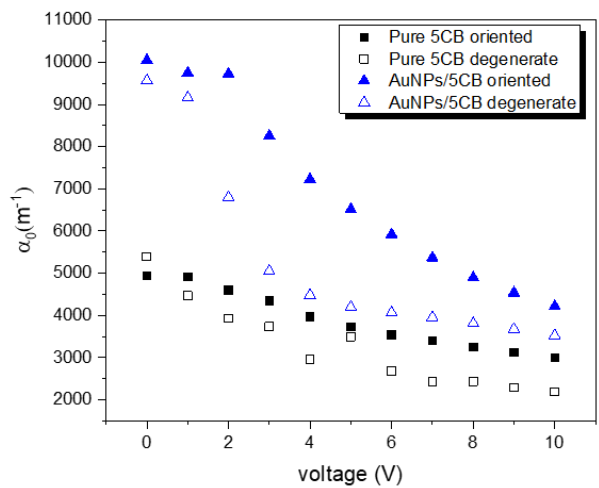

Figure 6 Linear absorption coefficients (a0) of all the samples at 532 nm under 1 kHz square wave voltages with peak-to-peak ranging from 0 to 10 V.

As the voltage is increased, the electric field induces reorientation of some of the LC molecules in the direction of the applied field. This causes a change in the birefringence of the LC layer, enabling some light to pass through the crossed polarizer-analyser configuration, as shown in Figures 5(b-d). The degree of light transmission corresponds to the extent of the director rotation caused by the electric field. As the voltage is further increased, more LC molecules reorient, leading to a decrease in light transmittance through the cell due to the perpendicular alignment of the LC director induced by the electric field. In the cross-polarized configuration study of a degenerate LC cell, V = 0 V will result in brighter samples, independently of the orientation of polarizers (Figure 5-e). This is because the L C molecules align in various directions allowed degenerate samples, the lack of a preferred anchoring direction makes the alignment of LCs molecules random, especially under varying external electric field, leading to a decrease in absorption as well. The linear absorption coefficients ($\alpha_0$) for all the samples at 532 nm is shown in figure 6. The values exhibit a decreasing trend as the voltage increases.

### 3.3 Nonlinear absorption

The third-order nonlinear optical characterization is performed by open-aperture Z-scan technique. All the samples reveal reverse saturable absorption behavior (Supplementary Information). By fitting the experimental curves to equation 1, the nonlinear absorption coefficient β of samples at different voltages were obtained and are plotted in figure 7. For pure 5CB, the values of β are very low and independent of the applied voltage across the LC cell, even though $\alpha_0$ is varied slightly with applied voltage. This exemplifies that the external electric field has no effect β while the orientation of the 5CB changes in response to the electric field, thereby altering the optical path and $\alpha_0$.

For AuNPs/5CB samples, β is maximized at 0 V and decreases with increasing voltage. Oriented sample (Vc = 1.5 V) shows the fastest decrease in β, reaching the value of pure 5CB, while degenerate sample (Vc = 7 V) show a slower rate of decrease. This behaviour is consistent with observations made using cross-polarized microscopy. It is important to note that this value of 1.5V pertains to the changes in the beta value and is not related to the Freedericksz transition. Specifically, 1.5V represents the point at which we begin to observe changes in beta. When a laser is incident on AuNPs, LSPR is excited along the polarization direction of the incident laser light. When no voltage is applied across the LC cell, the 5CB molecules align in the rubbing direction throughout the cell. As a result, dimers — pairs of interacting AuNPs — form along this rubbing direction, which is oriented perpendicular to the direction of light propagation and lies within the plane of the light's electric field amplitude. Thus, plasmon coupling resulted in increasing the nonlinear absorption coefficient, as depicted in figure 8(a). As the applied voltage increases, the electric field causes the 5CB molecules to reorient in the direction of the electric field, which is also the direction of light propagation. Since the AuNPs follow the orientation of the LC director, the formation of dimers will also align in the direction of propagation. This reorientation reduces plasmon coupling, which leads to a decrease in the nonlinear absorption coefficient, shown in figure 8(b). When the voltage exceeds a critical value, dimers are formed in the direction of the propagation of light, see figure 8(c), causing plasmon coupling to ceases completely.

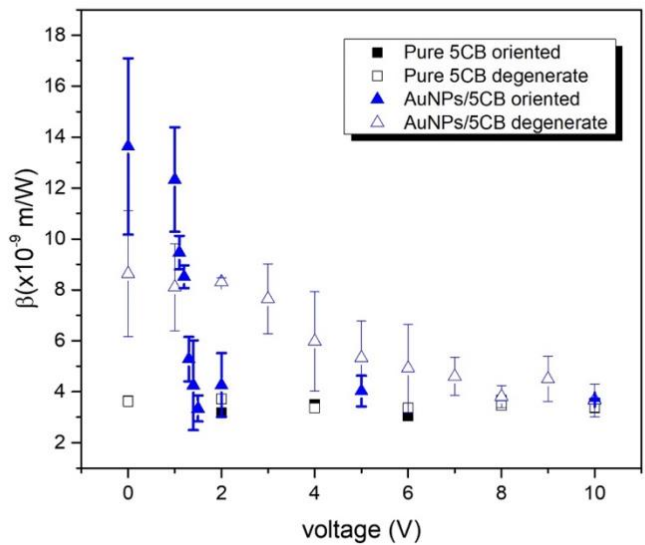

Figure 7 Third order nonlinear absorption coefficients (b) of all the samples at 532 nm under 1 kHz square wave voltages with peak-to-peak ranging from 0 to 10 V.

In degenerate samples, because of the absence of preferred alignment, the directors are randomly distributed throughout the sample. This randomness allows dimers to form in various directions, leading to plasmon coupling in multiple orientations. Furthermore, due to the presence of defects, which are more prevalent in random planar samples compared to uniform planar ones, can trap NPs. As a result, the NPs may not align with the electric field as they would in defect-free regions. As the applied voltage is increased, the electric field induces reorientation of the LC directors in the direction of the applied field. As the directors align with the electric field, the random distribution of dimers is disrupted, reducing the extent of plasmon coupling and, consequently β. However, due to the initial random distribution of the directors and the intermolecular coupling forces that resist reorientation, a higher voltage is required to achieve complete alignment of the LC molecules and cease the plasmon coupling.

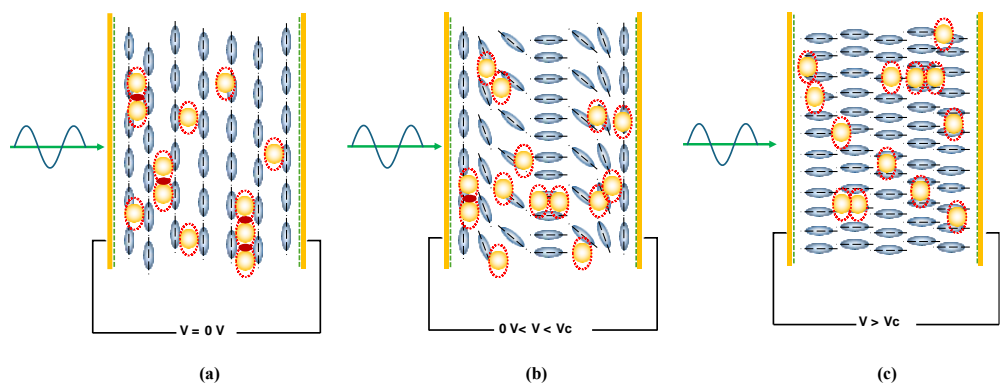

Figure 8 Schematic representation of the alignment and LSPR (dashed red line) of AuNPs within 5CB LC cells under varying electric field conditions: (a) is 0V, (b) below the critical voltage, and (c) above the critical voltage. The orientation of 5CB molecules in the LC cell, dictated by the rubbing direction, causes AuNP dimers to form perpendicularly to the light propagation direction, enhancing plasmon coupling (red ovals) and consequently increasing $\beta$.

We found that the variation of beta with voltage is consistent across different samples and locations within the same sample. Due to the size of the laser beam (2 mm unfocused and 65 microns focused), a significant number of nanoparticles (NPs) contribute to the observed nonlinearity, leading to an averaged response across the entire area. As a result, scanning various locations in the same sample yields consistent results, maintaining the same trend in beta variation with voltage.

To further understand the behavior of AuNPs samples with different anchoring conditions, the β values at different voltages were obtained utilizing polarization Z-scan technique. The procedure is similar to our previous study – using a linear polarizer and a quarter-wave plate the polarization of the incident beam is changed from 0˚ (polarization of light is in the rubbing direction), 45˚, and 90˚ (polarization perpendicular to the rubbing direction).[52] For the oriented AuNP/5CB sample (Figure 9 (a)), when the applied voltage is 0 V, β decreases as the polarization of light changes from 0˚ to 90˚, similar to our previous observation. When the applied voltage increases, the decreasing trend of β slows down and disappears when the voltage reaches the critical value. For degenerate sample (Figure 9 (b)), there is no such decreasing trend as the angle of polarization changes.

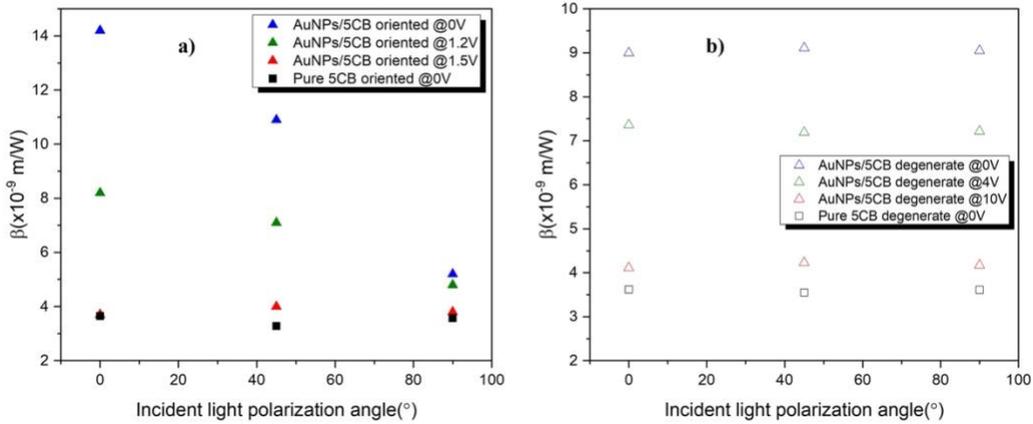

Figure 9 Variation of the β for different polarizations of light under various applied voltages, comparing oriented pure 5CB with a) planar oriented and b) degenerate AuNP/5CB.

Figure 10 shows comparison of $\alpha_0$ and $\beta$ before and after the voltage was applied demonstrates the robustness of the samples. The minimal difference in the coefficients before and after voltage application suggest that the samples exhibit reversible behaviour and do not experience significant damage or degradation, confirming that the applied voltage does not alter the intrinsic absorption properties of the samples. This stability under electrical influence is crucial for potential applications in optoelectronic devices where consistent optical properties are essential.

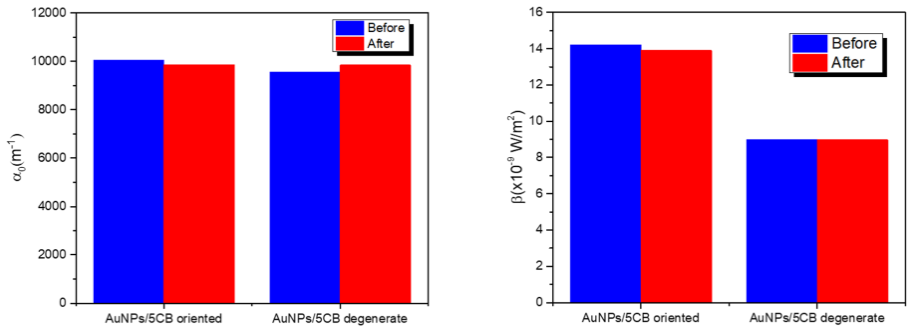

Figure 10 The comparison of (a) linear absorption coefficient and (b) nonlinear absorption coefficient of AuNPs/5CB samples before and after voltage applied. The stability of the absorption coefficients highlights the robustness and consistency of the samples.

## 4 Conclusion

In conclusion, our study highlights the dynamic tunability of AuNPs dispersed in 5CB nematic LCs for nonlinear optics applications. Through precise control of LC orientation using external electric fields, we effectively manipulate the plasmonic properties of AuNPs. In planar-oriented cells, the alignment of LC molecules and AuNPs perpendicular to incident light maximizes plasmon coupling and enhances third-order nonlinear absorption. Conversely, degenerate cells, with their lack of preferred alignment, show more varied responses but still exhibit controllable nonlinear optical effects. These results underscore the versatility of AuNP/5CB composites in developing adaptive photonic technologies, promising advancements in optoelectronics and beyond where responsive materials are crucial. Future research may explore optimizing nanoparticle configurations and LC parameters to further enhance these materials' performance and expand their applicability in dynamic optical systems.

# Supplementary information

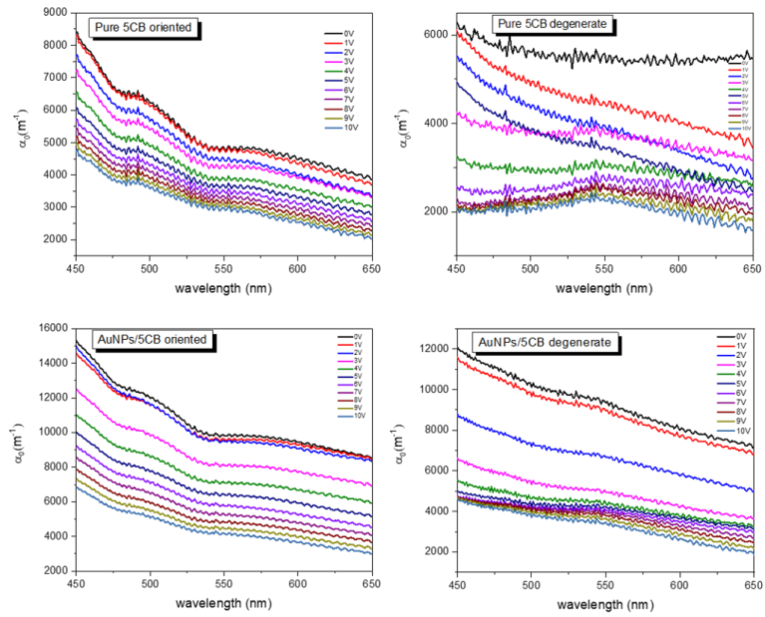

S1: Linear absorption coefficient spectra of all samples

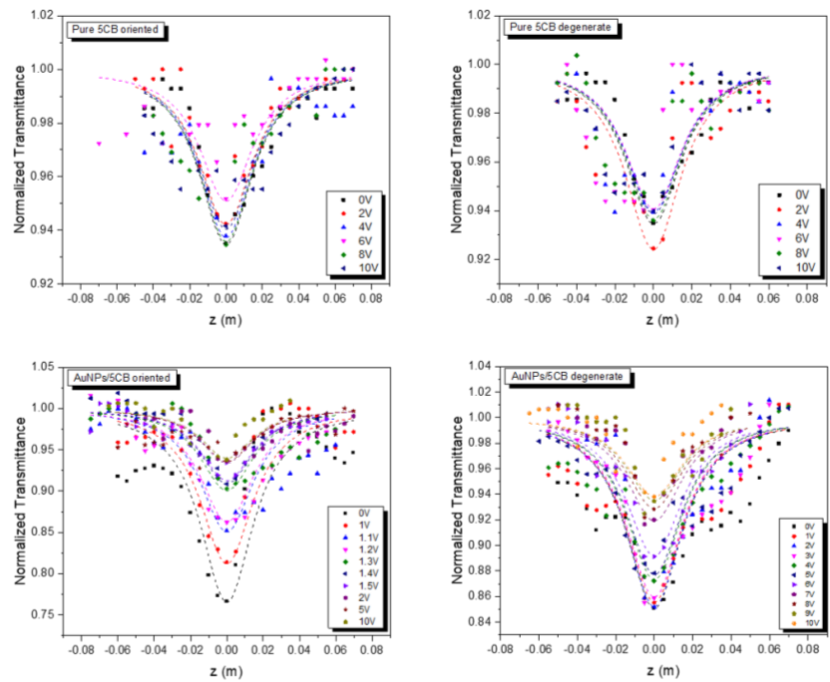

S2: Z-scan curves of all samples